\documentclass{aa}
\usepackage{graphicx}
\usepackage{amsmath,amssymb}
\usepackage{epsfig}                            
\usepackage{natbib}
\begin{document}
   \title{Localizing plages on 
          BO~Mic\thanks{Based on observations obtained at the ESO VLT}
         }

   \subtitle{Rapid variability and rotational modulation of stellar
   Ca~H\&K core emission}

   \author{U. Wolter
          \inst{1} 
          \and
          J.H.M.M. Schmitt \inst{1}
          }

   \offprints{U. Wolter, \\ \email{uwolter@hs.uni-hamburg.de}}

   \institute{Hamburger Sternwarte, Gojenbergsweg 112, D-21029 Hamburg, Germany \\
              \email{uwolter@hs.uni-hamburg.de, jschmitt@hs.uni-hamburg.de}
             }

   \date{Received / Accepted}

   \abstract{
We have obtained a densely sampled time series of Ca\,\textsc{ii}~H\&K
line profiles of the ultrafast rotating \mbox{K-dwarf} star \object{BO~Mic}.
Taken at high resolution, the spectra
reveal pronounced variations of the emission core profiles.
We interpret these variations as signs of
concentrated chromospherically active regions, in analogy to solar plages.
We further interpret the
variations as partly due to the rapid growth and decay
of plages, while other variations appear to be caused
by plages moved over the visible stellar disk by rotation.
The equivalent width of the Ca~K core emission changes
approximately in anti-phase to the photospheric brightness, suggesting
an association of the chromospheric plage regions with
pronounced dark photospheric spots.
We believe that further analysis of the presented spectral time series
will lead to a chromospheric Doppler image of BO~mic.

   }

   \maketitle
%
\section{Introduction}
\vspace*{-1.0\medskipamount}
The  core emission of the Ca\,\textsc{ii}~H\&K lines (3968.49~\AA\
and~3933.68~\AA) has become an indispensable tool of stellar atmospheric
diagnostics, starting with the pioneering observations of 
\citet{Eberhard13}.
The cores of Ca~H\&K are formed in high layers of the stellar atmosphere,
they react sensitively to
the temperature there, i.e. to chromospheric activity.

For the Sun, three components can be discerned in spatially resolved images of the chromosphere:
(i)~Quiescent regions;
(ii)~the chromospheric network
covering the whole solar surface;
(iii)~extended regions with increased chromospheric emission, 
associated with active regions, called plages. 
Whether a similar hierarchy of chromospherically active
regions also exists on stars other than the Sun is only poorly known
at present. The observed variability of stellar Ca~H\&K core emission is
usually interpreted as a superposition of rotational modulation and
intrinsic long-term variations \citep[e.g.][]{Vaughan81, Char93}. 
This interpretation suggests a
concentration of chromospheric activity in plage-like regions,
leading to the rotational modulation, embedded in more homogeneously
distributed chromospherically active regions producing the observed
basal fluxes.  

Previous studies of Ca~H\&K emission
concentrated on \textit{either} long-term variability observed at
relatively low spectral
resolution \citep[e.g.][]{Baliunas98} \textit{or} high-resolution ``snapshots''
\citep[e.g.][]{Linsky79, Pasquini88}.
Only a few studies combine regular phase sampling with high or
moderate spectral resolution in order to attempt localizing chromospheric
features. \citet{Neff89} and \citet{Busa99} 
analyzed UV observations of the Mg\,\textsc{ii}~h\&k lines 
carried out by the IUE satellite. By applying
multi-Gauss fits to the rotationally broadened line
profiles of the RS~CVn systems \object{HR~1099} and \object{AR~Lac}, these authors 
could detect weak rotational modulation of the profiles and 
localize chromospheric emission regions on the stellar surface.
Because of the barely resolved rotational line broadening
and limitations due to the noise level  of
the spectra, only rough localizations were possible.

In the following, we present 
first results concerning a more precise localization of
chromospheric emission features on the surface of the ultrafast
rotating, apparently single young K-dwarf star BO~Mic, usually
nicknamed \mbox{``Speedy Mic''}
(K2V, 
\mbox{$P_{\mathsf{rot}}=0.380\pm0.004$~days}).
We use observations combining high spectral
resolution, dense and continuous phase sampling and high SNR.  
Its fast rotation and high level of activity \citep[e.g.][]{Bromage92} make BO~Mic a
formidable object for such a study. 

\vspace*{-1.3\medskipamount}
\section{Observations}
\vspace*{-0.85\medskipamount}
The observations were performed with the VLT at the ESO Paranal on two nights (2002 August~2 and~7),
continuously sampling two complete rotations of BO~Mic.
The spectrograph \textsc{Uves} was used in a dichroic mode,
resulting in covered spectral ranges of 3260~\AA\ to 4450~\AA \ and 4760~\AA\ to 6840~\AA\  
at a spectral resolution of $\lambda / \Delta\lambda\approx40\,000$.

The red arm spectra have been analyzed to produce high-resolution
Doppler images of photospheric spots on BO~Mic's surface
(Wolter et al. 2005, below \citealt{Wolter05}), details of the data reduction can be found there.
In total, we obtained 273~spectra of BO~Mic;
these spectra have been added in pairs for the subsequent analysis,
discarding 18~spectra 
because of poor SNR due to clouds.
The resulting spectra have typical exposure times of about 450~s;
their average separation is 540~s. 
The SNR range from 60 to 120, with a typical value of 90.
The following discussion concentrates on the
August~2 spectra;
the August~7 spectra are of equal quality and show a qualitatively very
similar behaviour.

\vspace*{-1.0\bigskipamount}
\section{Results}
\vspace*{-0.85\medskipamount}
\subsection{Rotational broadening}
\label{sec:rot-broa}
\vspace*{-0.85\medskipamount}
\begin{figure}
 \center{
  \epsfig{file=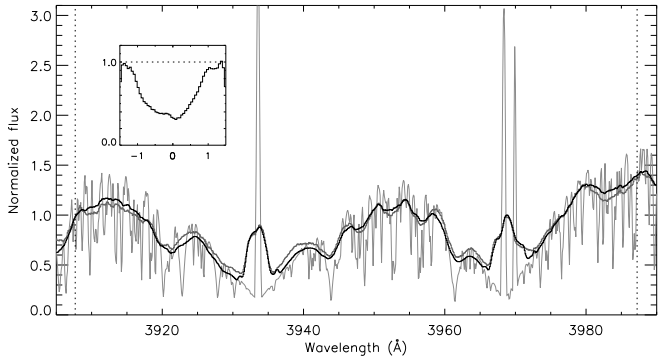,
  width=\linewidth,clip=, bb= 40 375 540 650}
        }
  \vspace*{-\bigskipamount}
  \caption{
    A spectrum of BO~Mic 
    (rotation phase 0.306, gray
    broad-lined spectrum),
    approximately fitted by a
    rotationally broadened template spectrum.
    The fit is rendered in black, 
    the broadening function is shown in the
    small inset panel. 
    The narrow-lined template spectrum  was generated from an observed K-dwarf
    spectrum by massively amplifying its Ca~H, Ca~K and
    H$_\epsilon$ emission cores. See text for details.
          }
  \label{fig:SpMic+HD155885mod} 
\end{figure}
Fig.~\ref{fig:SpMic+HD155885mod} shows one spectrum from the recorded time series.
It is
compared to a fit obtained by convolving a narrow-lined
template spectrum with a broadening function. The broadening function
models the Doppler-broadening due to rotation and the distribution
of emission regions on the stellar surface.  
The fit approximately reproduces all features of the BO~Mic spectrum;
this illustrates that the Ca~H\&K emission cores 
of BO~Mic can be interpreted as resulting from emission regions
distributed over essentially the whole stellar surface.
The broadening function was determined by our deconvolution procedure
sLSD (``selective least squares deconvolution'', cf. \citealt{Wolter05}).

The used template spectrum has been generated in the following way:
We started out from the observed spectrum 
of a chromospherically active star of similar spectral type 
with a low projected rotational velocity $v\,\sin{i} \le 10$~km/s 
(\object{HD~155885}, K1~V). 
In order to crudely simulate a star with much stronger chromospheric activity,
we modified HD~155885's spectrum by
amplifying the  emission
cores of both the Ca~H\&K lines 
by a factor of~5. Additionally, we added a tentative emission core to
the H$_\epsilon$ line (3970.07~\AA).
Only adding these amplified line emission cores leads to a
qualitatively  correct fit to the observed spectra of BO~Mic.


\vspace*{-2.0\medskipamount}
\subsection{Emission core variations}
\vspace*{-0.85\medskipamount}
\begin{figure}
 \center{
  \epsfig{file=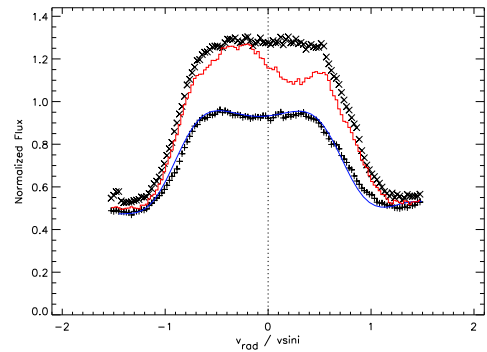, width=0.78\linewidth,clip=}
        }
  \caption{
    Variations of the Ca~K line core observed for
    BO~Mic during two stellar rotations.
    The symbols show the maximum~({\tiny $\times$}) and
    minimum~({\tiny $+$}) flux observed at each radial velocity $v_\mathsf{rad}$; 
    note that neither the minimum nor the maximum profile have been observed
    at a single phase.
    The upper red/gray line shows one sample profile from the time series
    with a typical SNR
    (phase 0.616). The lower blue/gray profile shows a fit to the minimum
    profile 
    representing a uniform surface distribution of emission features, subjected to
    a linear \textit{limb brightening}. 
          }
  \label{fig:SpMic_CaKcore_minmax} 
\end{figure}
\begin{figure*}
 \center{
  \epsfig{file=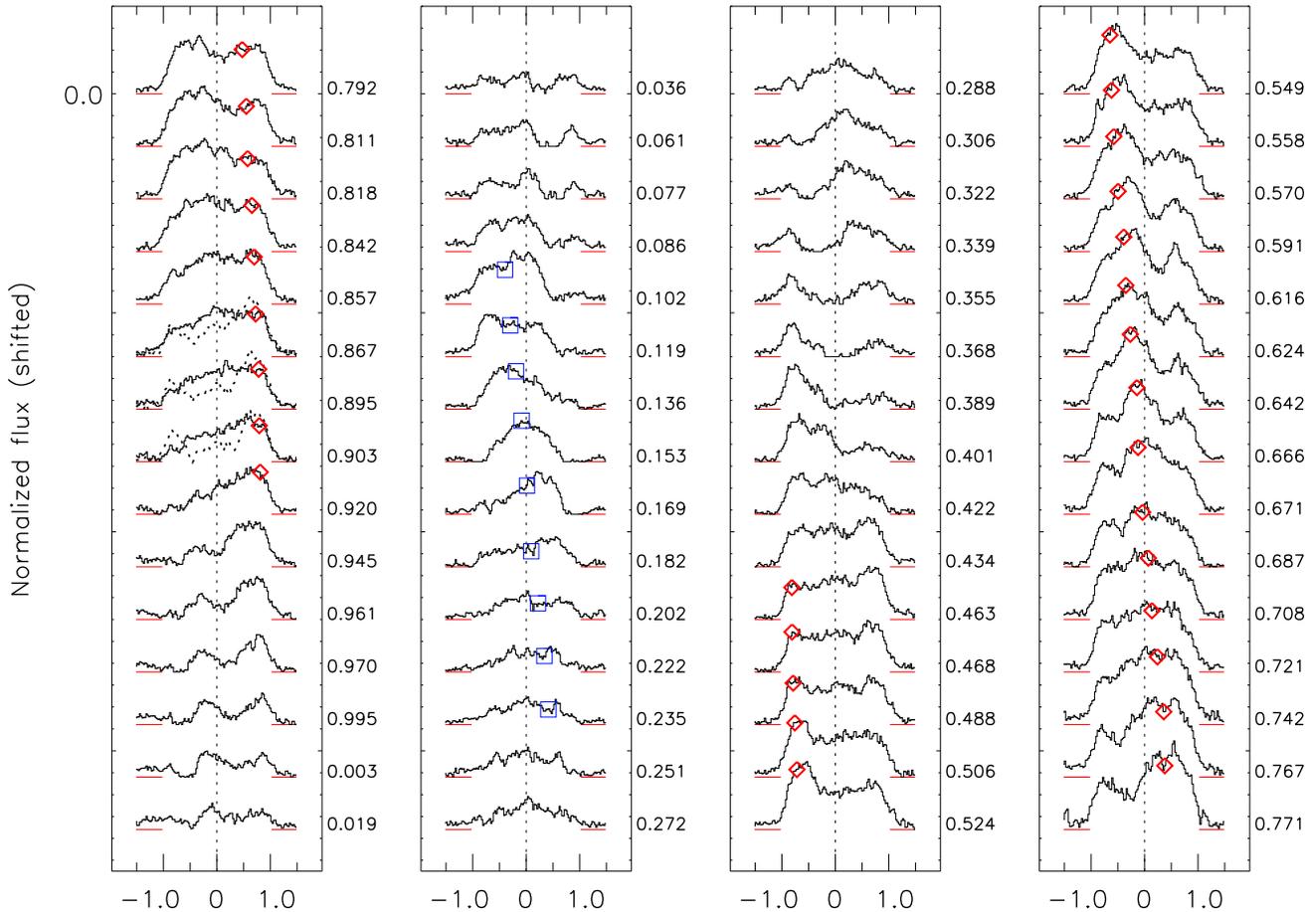, width=0.99\linewidth,clip=, bb= 30 285 580 665}
        }
  \caption{
    Residual Ca~K core profile variations of BO~Mic.
    The continuous profiles cover one complete stellar
    rotation, observed on 2002
    August~2; the dotted lines show three sample profiles observed on August~7 
    for comparison.
    The rotation phase is given right of each graph which
    shows the difference of the observed profile to the minimum
    profile plotted in Fig.~\ref{fig:SpMic_CaKcore_minmax}. 
    Subsequent profiles are shifted by 0.24 in flux.
    The symbols mark the radial velocity of tentative
    features fixed on the stellar surface.
    Note that
    the squares are \textit{not} associated with any
    discernible migrating deformations of the profiles,
    see text.
          }
  \label{fig:SpMic_CaKcores_aug2} 
\end{figure*}
Fig.~\ref{fig:SpMic_CaKcore_minmax} provides a more detailed view of the
Ca~K core profile of BO~Mic. To illustrate the total range of
observed variations, the upper and lower envelope of all
observed profiles have been plotted; these are termed the
maximum and minimum profile in the following, respectively.

The line profiles have been normalized to the surrounding line flanks
of the Ca~K line (between 3923.5~\AA\ and 3925.5~\AA).
This normalization introduces no significant error when comparing Ca~K cores at
different phases: As the maximum and minimum profiles in
Fig.~\ref{fig:SpMic_CaKcore_minmax} show, the line flanks
outside the emission core do not show variations above the noise level.
The wavelength scale has been transformed to units of projected
rotation velocity, using a value of \mbox{$v\,\sin{i}=$ 134~km/s},
as determined by \citet{Wolter05}.

Fig.~\ref{fig:SpMic_CaKcore_minmax} clearly illustrates the large amplitude of the variations.
Practically all core profiles show pronounced deformations, 
making most of them significantly asymmetric.
In contrast, the symmetry of the maximum and minimum
profiles is quite striking.

As also shown in Fig.~\ref{fig:SpMic_CaKcore_minmax}, this minimum
profile can be closely approximated by convolving the template
spectrum of Fig.~\ref{fig:SpMic+HD155885mod} 
with an analytic
rotation profile \citep{Gray92, Wolter05}. 
More precisely, the Ca~K core of the template
spectrum has been reduced by
80\% in equivalent width and broadened by
a Gaussian of 0.8~\AA\ width in order to obtain a close fit to the minimum 
profile flanks.
This convolution describes the broadening of the line core due to
rotation, assuming emission regions homogeneously
distributed over the whole stellar surface.
It adopts a linear limb
darkening law
with a
limb darkening parameter 
of
\mbox{$\epsilon = -5$}, i.e. it represents a \textit{limb brightening}.
Regardless of the adopted linear law,
a pronounced limb brightening is required to approximate the line core profile.

\begin{figure}
 \center{
  \epsfig{file=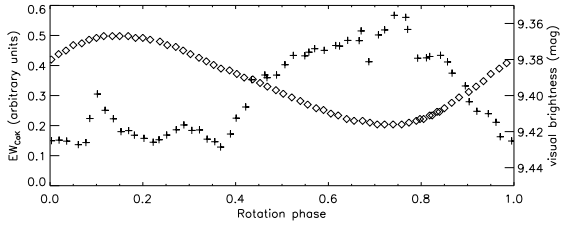, width=0.85\linewidth,clip=, bb= 65 290 560 500}   
        }
  \caption{
    Equivalent width variations $EW_{\mathsf{CaK}}$ of the CaK core emission ({\tiny $+$}) and the
    photospheric brightness ($\diamond$) of BO~Mic on 2002 August 2,
    as a function of rotation phase. Lacking strictly
    simultaneous photometry to our spectral observations, the
    photospheric brightness has been calculated from our Doppler images
    \citep{Wolter05}. A value of 
    \mbox{$EW_{\mathsf{CaK}}=0$} would correspond to the minimum
    profile shown in Fig.~\ref{fig:SpMic_CaKcore_minmax}. 
           }
  \label{fig:sm_EWcaK-6400lc} 
\end{figure}

The variations of the 
core equivalent width are plotted in Fig.~\ref{fig:sm_EWcaK-6400lc}.
They should be interpreted with some care, since they depend on the
level of chromospheric emission \textit{and} the value of the
underlying photospheric continuum. A comparison with the
simultaneous photospheric lightcurve plotted in the same figure
(reconstructed from our Doppler images, cf. \citealt{Wolter05})
shows that the maximum core equivalent width 
indeed coincides closely with the photospheric brightness
minimum.
Qualitatively, this would even result
from a constant chromospheric emission level, due to photospheric
variations alone. However, the Ca~K core flux variations are far too strong to be explained
in this way. They amount to about $\pm20\%$
of the average flux (cf. Fig.~\ref{fig:SpMic_CaKcore_minmax}), while the photospheric 
brightness varies only by about $\pm2\%$.
However, the \textit{shape} variations of the Ca~K core are not affected by
this argument;
only these shape variations are analyzed in the following.

\vspace*{-2.0\medskipamount}
\subsection{Rotational modulation}
\vspace*{-0.85\medskipamount}
Fig.~\ref{fig:SpMic_CaKcores_aug2} shows the continuously sampled variations of the Ca~K
line core for one complete rotation of BO~Mic. 
Rotation phases $\phi$ were calculated using the same ephemeris as
\citet{Wolter05}, based on the rotation period determined by \citet{Cutispoto97}:
  \mbox{$HJD = 2\,448\,000.05 + 0.380 \cdot \phi$}.\
For each rotation phase the difference of the observed and the minimum profile
has been plotted. 
Here, this procedure
merely serves a compact graphical representation; 
the variations are equally well  visible in the observed core profile
without this subtraction (cf. Fig.~\ref{fig:SpMic_CaKcore_minmax}).

Many of the core shape deformations change too fast to be attributable to
rotational modulation.
An example is illustrated in Fig.~\ref{fig:SpMic_CaKcores_aug2} 
by the square-marked feature at phases~0.102~-~0.235. 
The squares mark the rotationally induced migration of a profile
deformation caused by a tentative equatorial feature. 
Such a profile deformation shows the fastest possible migration
associated with a fixed feature directly on the stellar surface. 
Note that the
squares do not match the observed evolution of the most pronounced
deformations at these phases. 
As a consequence these variations must be caused by intrinsic
variations of the emitting regions on the surface,
i.e. by their growth or decay. Note that some of these changes take place on
timescales of about 10~minutes, the typical separation of two
subsequent profiles. 

On the other hand, several variations of the core shape
are clearly suggestive of rotational modulation:
They show the characteristic appearance in the blue wing of the line
core and a migration towards the red. As an example, the diamonds
in Fig.~\ref{fig:SpMic_CaKcores_aug2} approximately mark one such
``bump'' apparently migrating though the profiles.
The corresponding surface feature would be located at mid-latitude on
the stellar surface (at $35\degr$\ latitude and $110\degr$\ longitude,
using the coordinates of \citealt{Wolter05}).

\vspace*{-1.1\bigskipamount}
\section{Summary and discussion}
\vspace*{-0.85\medskipamount}
We study the Ca~H\&K emission cores of the
ultrafast rotating pre-main-sequence
star BO~Mic,
our observations continuously cover two complete stellar rotations. 
The observed line cores 
can be approximated by the rotationally broadened template spectrum of 
a slowly rotating star.
The used template spectrum models an 
extremely chromospherically active star;
it was generated from an observed
K-dwarf 
spectrum by massively amplifying the emission
cores of the Ca~H\&K and H$_\epsilon$ lines.
No slowly rotating late-type star known to us shows a level
of chromospheric emission comparable to the resulting template spectrum, 
confirming BO~Mic's strong
chromospheric activity.

Despite the large asymmetry of the individual emission profiles,
the minimum Ca~K core profile reconstructed from all observed spectra
is nearly symmetric, it can be closely approximated using an
analytic rotational broadening function. This approximation suggests
to interpret the minimum profile as a proxy of BO~Mic's ``basal''
chromospheric emission, originating from regions homogeneously distributed
over the whole stellar surface. 
In this respect, these regions are possible analogues of the solar chromospheric
network; this interpretation is supported by the inferred \textit{limb
brightening} which is also observed for solar chromospheric
faculae when measuring their contrast to the photospheric continuum.


The equivalent width of the Ca~K core varies approximately in
anti-phase to the photospheric brightness.
This strongly suggests that plage-like chromospheric 
emission regions 
on BO~Mic's surface are associated with active regions, hence with dark spots.

The observed Ca~H\&K line cores show pronounced and
  strongly variable deformations.  Due to the
  large  $v\,\sin{i}$ of BO~Mic, these deformations can be assigned quite precise
  radial velocities in the Doppler-broadened profiles. 
  As a result, they can be associated with well-defined emission regions on the
  stellar surface, assuming that chromospheric flows are
  slow compared to the rotational velocities.

The line cores vary
significantly on timescales down to about 10~minutes.
The manifold core
profiles indicate a
complex distribution of chromospherically active regions
  on the surface of BO~Mic subjected to rapid intrinsic evolution.
  Several deformations of the time series show a migration
  through the line profile consistent with rotational
  modulation, i.e. with emission regions fixed on the
  surface, moved over the visible stellar disk by rotation.

Using the  Ca~K core profile deformations to localize strong chromospheric emission regions
on BO~Mic's surface 
does not lead to a straightforward
identification with pronounced features of our photospheric Doppler images.
We believe that this is due to the complex surface distribution of the emission 
regions, making a Doppler imaging analysis of the Ca~H\&K line profiles necessary.
We are confident that a
further analysis of our observations 
will lead to a chromospheric Doppler image of BO~Mic.

\vspace*{-\smallskipamount}
\begin{acknowledgements}
      U.W. acknowledges financial support from
      \emph{Deut\-sche For\-schungs\-ge\-mein\-schaft}, \mbox{DFG -
      SCHM 1032/21-1}.
\end{acknowledgements}

\end{document}